\documentclass[aps,twocolumn,prl,tightenlines,floatfix,showpacs]{revtex4}
\usepackage[dvips]{graphicx}
\usepackage[english]{babel}  
\usepackage{amsmath}  
\usepackage{amssymb}  
\usepackage{slashed}  
\usepackage{feynmf}
\usepackage{marginnote}

\begin{document}

\title{Spin Transport in Cold Fermi gases: A Pseudogap Interpretation
of Spin Diffusion Experiments at Unitarity}

\author{Dan Wulin $^{1}$, Hao Guo $^{2}$,
Chih-Chun Chien$^{3}$ and K. Levin$^{1}$}

\affiliation{$^1$James Franck Institute and Department of Physics,
University of Chicago, Chicago, Illinois 60637, USA}

\affiliation{$^2$Department of Physics, University of Hong Kong,
Hong Kong, China}

\affiliation{$^3$Theoretical Division, Los Alamos National Laboratory, MS B213, Los Alamos, NM 87545, USA}

\date{\today}

\begin{abstract} 
We address recent spin transport experiments in ultracold unitary Fermi gases.
We provide a theoretical understanding for
how the measured temperature dependence of the spin diffusivity
at low $T$ can disagree with the expected behavior of a
Fermi liquid (FL) while the spin susceptiblity
(following the experimental protocols) is consistent
with a Fermi liquid picture.
We show
that
the experimental
protocols for extracting $\chi_s$ 
are based on a FL presumption; relaxing this leads
to consistency within (but not proof of) a pseudogap-based
approach. 
Our tranport calculations
yield insight into the measured strong
suppression of the spin diffusion constant at lower $T$.
\end{abstract}

\maketitle

Recent measurements 
associated with mass transport \cite{ThomasViscosityScience_online}
and spin transport \cite{ZwierleinSpin} 
in the ultracold Fermi
gases and superfluids are of great interest principally because they
provide detailed information about the excitation spectrum,
thereby strongly constraining microscopic theories. Equally important
are their widespread implications for a host of different
strongly correlated systems ranging from quark-gluon
plasmas to the high $T_c$ cuprates \cite{NJOP}.
Common to these materials are the short mean free paths
which lead to ``near-perfect fluidity"
\cite{ThomasViscosityScience_online}
``bad metallicity" \cite{Emery} and now bad spin conductivity 
\cite{ZwierleinSpin}.
These latter are Fermi-gas-based
materials with small spin diffusivities $D_s$ that approach the 
quantum limited value $\hbar/m$ as the temperature is lowered. 
However, the nature of the excitations in these Fermi gases is
currently under debate \cite{SalomonFL,Salomon4} and
a controversy has emerged as to whether
the associated normal state of these superfluids
is a Fermi liquid or contains an excitation (pseudo)gap.
Moreover, there are mixed features in recent spin transport
experiments \cite{ZwierleinSpin} which are indicative of both pairing and Fermi liquid
theory.

The goal of this paper is to address these  
spin transport experiments by Sommer et al. \cite{ZwierleinSpin} which measure
the spin diffusivity $D_s$, and the spin conductivity $\sigma_s$
and thereby deduce the spin 
susceptibility $\chi_s = \sigma_s/D_s$,
as functions of temperature at 
unitarity. 
Our calculations, based on
a pseudogap approach to BCS-BEC crossover theory \cite{Ourreview}, 
have the aim of resolving apparent 
experimental descrepancies between interpretations
of $\chi_s$, $D_s$ and $\sigma_s$.
It should be stressed, however, that we focus on lower $T$ 
(below the pairing onset $T^*$) than
in Ref.~\onlinecite{ZwierleinSpin} where one can fully contrast Fermi
liquid and pseudogap theories.
Indeed, if one is to sort out whether the normal state of a unitary gas
has a pseudogap or is a Fermi liquid, it is essential to address
these spin (as well as mass) transport experiments using
multiple theoretical frameworks.

Our key observation is that one must differentiate between the spin
susceptibility 
as computed assuming a constant number of
carriers in $\sigma_s$ and one which is
calculated assuming the number of carriers in $\sigma_s$ increases with $T$
reflecting the presence of a pairing gap.
For the latter case
one recovers a consistent interpretation of
both $\chi_s$ and $D_s$, in contrast to the former Fermi liquid
approach, which leads to this mixed interpretation.

The physics of transport in the unitary gas is complicated
because, if a pseudogap is present in the
normal state, one should accomodate both
fermions and fermionic pairs.
In recent papers we have included both types of quasi-particles
in addressing \cite{Ourviscosity} the observed anomalously low
shear viscosity 
\cite{ThomasViscosityScience_online}
both above and below $T_c$, 
as well as magnetic and non-magnetic Bragg scattering \cite{OurBraggPRL}.
While previous studies of spin transport in the cold Fermi gases have 
employed,  
for example, Quantum Monte Carlo, Boltzmann transport and variational approaches 
\cite{HuseSpin,BruunSpin},
we choose to use
the Kubo formalism which more readily addresses
conservation laws and sum rules.
The underlying theory is a BCS-BEC crossover 
theory in a t-matrix approximation \cite{Ourreview}. At $T=0$, the 
system is in the BCS-Leggett ground state where there are only superconducting pairs 
(characterized by the gap $\Delta_{sc}$). For temperatures $0<T\leq T_c$, 
there is a comixture of superconducting pairs, noncondensed pairs 
(characterized by the gap $\Delta_{pg}$), and fermionic
quasiparticle excitations. 

\begin{figure*}
\includegraphics[width=2.1in,clip]
{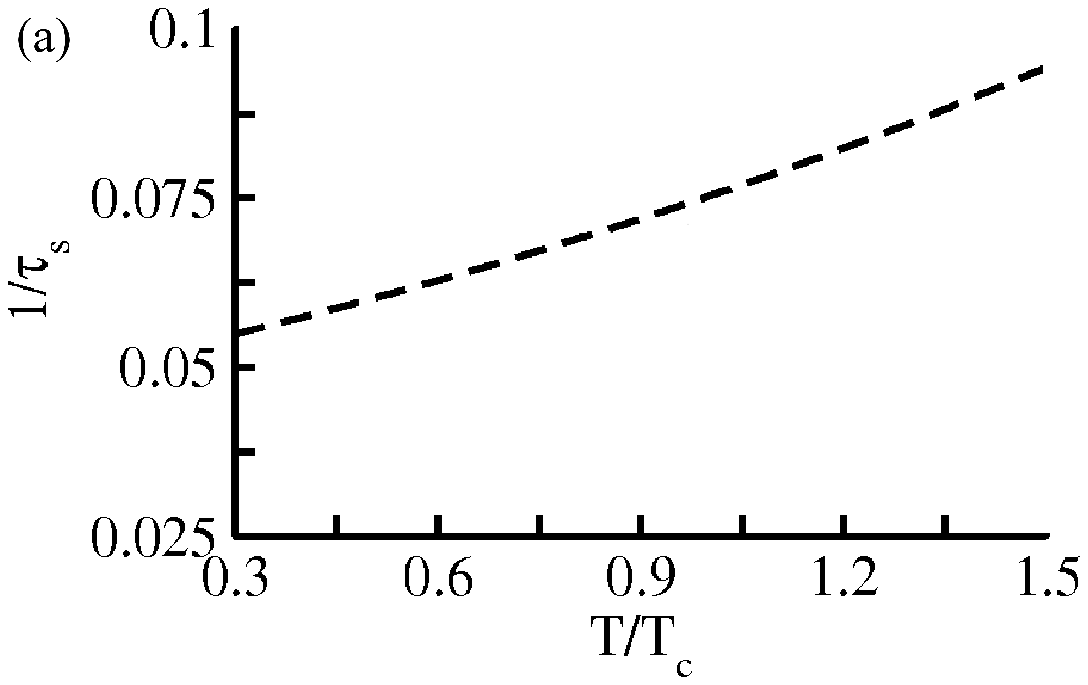}
\includegraphics[width=2.1in,clip]
{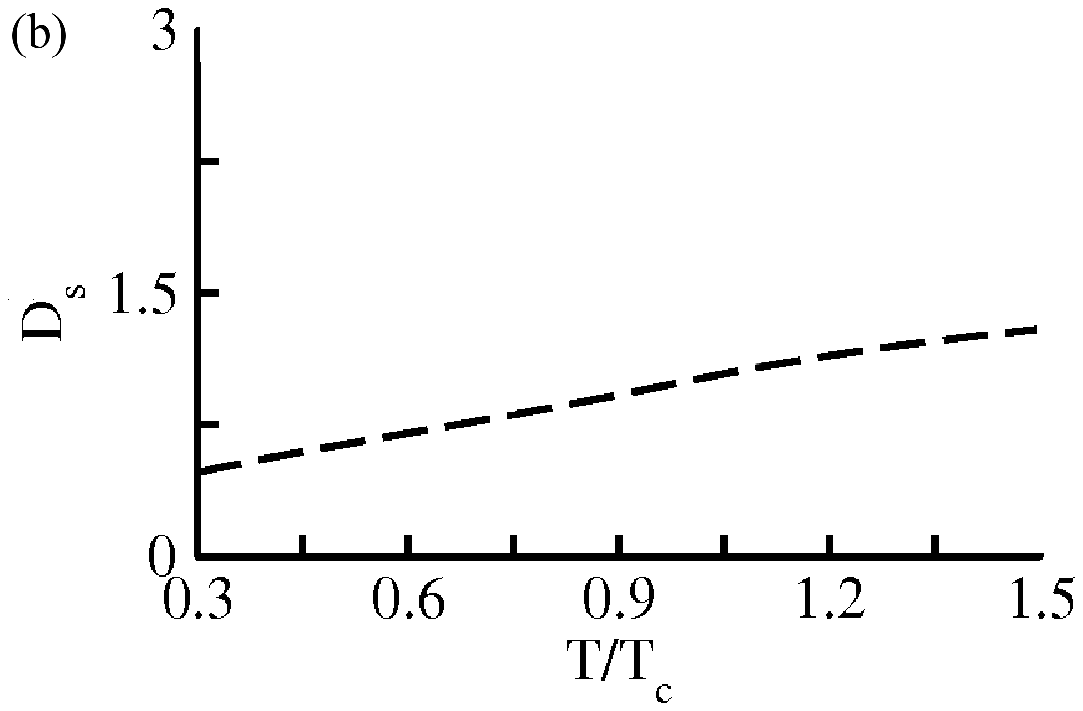}
\includegraphics[width=2.1in,clip]
{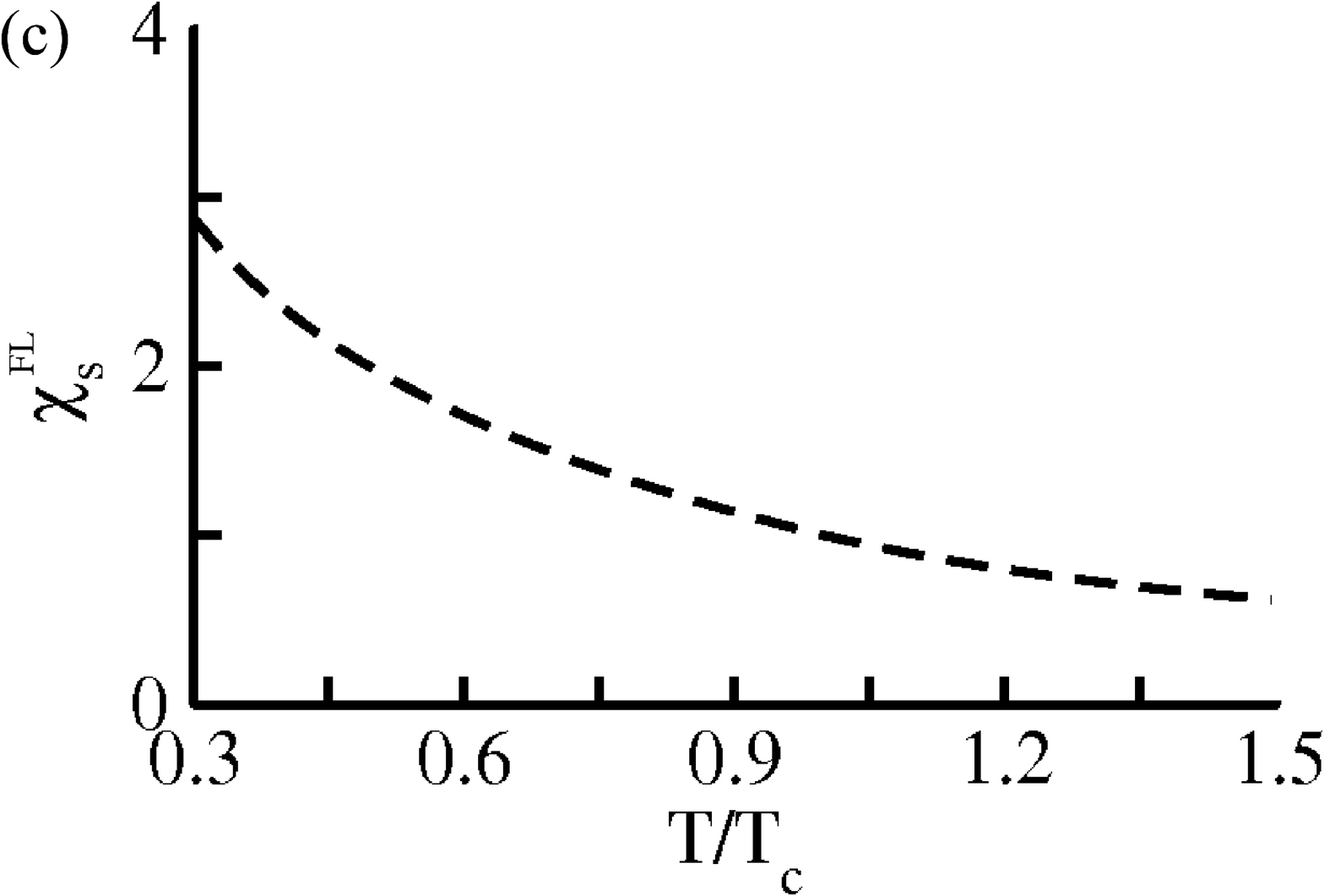}
\caption{
(a) The quasiparticle inverse lifetime $\tau_{s}^{-1}$ in units $E_f$. The values are inferred from RF experiments \cite{OurComparison}
and are used as input in our microscopic calculations \cite{Ourviscosity}. The temperature range is limited to $0.3T_c<T<1.5T_c$ where the RF-derived lifetime data are available. 
(b) The spin diffusivity $D_s$ for a homogenous unitary Fermi gas in our microscopic
 theory. It is normalized by its value at $T_c$. 
(c) The spin susceptibility $\chi_s^{FL}$
computed assuming a temperature independent effective number of carriers and normalized by the value at $T_c$. 
At low $T$, Figs. (a), (b) and (c) can be directly compared to Figs. 2,3 and 4a 
respectively in Ref. \onlinecite{ZwierleinSpin}.}
\label{fig:1}
\end{figure*}

Essential to satisfying conservation laws and sum rules
is the incorporation of Ward identities
and collective mode physics. (The latter applies only
to the mass transport or ``electromagnetic" response.)
These issues are addressed in Refs.~\onlinecite{Kosztin2,OurBraggPRL}.
The presence of both condensed and
non- condensed pairs in transport  
leads to
the usual Maki-Thompson (MT) and Aslamazov-Larkin (AL)
diagrams, which are demonstrably
consistent with gauge invariance. 
A different diagram sub-set necessarily appears in
the spin response, as compared with the mass-transport
(or electromagnetic response); in the former the
AL
diagrams are not present, nor do collective mode (or phononic
Goldstone bosons) enter \cite{OurBraggPRL,Combescot06}.
The fermionic self energy
on which our work is based has become the \textit{literature standard} for
($ T > T_c$) pseudogap theories \cite{SentLee,Levchenko}
\begin{eqnarray}
\Sigma(\textbf{p},i\omega_n)=-i\gamma +\frac{\Delta_{pg}^2}{i\omega_n+\xi_{\textbf{p}}+i\gamma} +\frac{\Delta_{sc}^2}{i\omega_n+\xi_{\textbf{p}}}
\label{eq:1}
\end{eqnarray}
where $\xi_{\textbf{p}}$ is the free quasiparticle dispersion and $\gamma$
is the fermionic inverse-lifetime associated with the conversion
from fermions to non-condensed pairs.
This 
general form for $\Sigma$ in Eq(\ref{eq:1}) has been applied in both experimental\cite{Jin6} and theoretical\cite{RFReview} radio frequency (RF) 
studies of the cold Fermi gases. 
In the weak dissipation limit where $\gamma$ is small,
there is little distinction between condensed and non-condensed pairs,
while in
the strong dissipation limit, this distinction is enforced.
In this
paper we find that these two approaches tend to converge for
$s$-wave pairing and the lifetime parameters obtained from RF experiments
\cite{OurComparison}.
For reasons of transparency, for the most part
the equations we present are
in the weak dissipation limit.

We write the key equations first 
for the dc spin conductivity
in the more general strong dissipation limit
and second 
for the
the spin-spin correlation function at general $(\bf q , \omega)$ (called 
$Q_{00}^s (\mathbf{q}, \omega)$)
but in the limit in which the non-condensed pair lifetime is
relatively long: 
\begin{widetext}
\begin{eqnarray}
\sigma_s&\!\!=\!\!&-\lim_{\omega\rightarrow0}\lim_{q\rightarrow0}\frac{1}
{6m^2\omega}\textrm{Im}\sum_P\textbf{p}^2
\Big[G_{P^+}G_{P}-F_{\textrm{sc},P^+}F_{\textrm{sc},P}
-F_{\textrm{pg},P^+}F_{\textrm{pg},P}\Big]_{i\Omega_l\rightarrow\omega^+},
\label{eq:40}
\end{eqnarray}

\begin{eqnarray}
Q_{00}^s (\mathbf{q},\omega)
=
\sum_{\mathbf{p}}\Big[\frac{E^+_{\mathbf{p}}
+E^-_{\mathbf{p}}}{E^+_{\mathbf{p}}E^-_{\mathbf{p}}}\frac{E^+_{\mathbf{p}}
E^-_{\mathbf{p}}-\xi^+_{\mathbf{p}}\xi^-_{\mathbf{p}} - \Delta^2_{\textrm{sc}}
- \Delta^2_{\textrm{pg}}}{\omega^2-(E^+_{\mathbf{p}}+E^-_{\mathbf{p}})^2}
\big(1-f(E^+_{\mathbf{p}})-f(E^-_{\mathbf{p}})\big)\nonumber\\
\quad-\frac{E^+_{\mathbf{p}}-E^-_{\mathbf{p}}}{E^+_{\mathbf{p}}E^-_{\mathbf{p}}}
\frac{E^+_{\mathbf{p}}E^-_{\mathbf{p}}+\xi^+_{\mathbf{p}}\xi^-_{\mathbf{p}}
+ \Delta^2_{\textrm{sc}} + \Delta^2_{\textrm{pg}}}{\omega^2-(E^+_{\mathbf{p}}-
E^-_{\mathbf{p}})^2}\big(f(E^+_{\mathbf{p}})-f(E^-_{\mathbf{p}})\big)\Big],
\label{eq:17}
\end{eqnarray}
\end{widetext}
where $P^+=(\mathbf{p}+{\mathbf{q}},i\omega_n+i\Omega_l)$,
$P=(\mathbf{p},i\omega_n)$, and $\omega^+=\omega+i0^+$. The quantity $i\omega_n$ ($i\Omega_l$) is a fermionic (bosonic) Matsubara frequency. The excitation energy is $E_{\textbf{p}}=\sqrt{\xi_{\textbf{p}}^2+\Delta^2}$ 
where
$\Delta=\sqrt{\Delta_{sc}^2+\Delta_{pg}^2}$. The supercripts $\pm$ on $\xi^{\pm}$ and $E^{\pm}$ indicate the momentum argument $\textbf{p}\pm\frac{\textbf{q}}{2}$. 
Here
$f$ is the Fermi function and
$G= G(\Sigma)$ is the dressed Green's function.
Note that while one can interpret $F_{sc}$ as the usual Gor'kov Greens
function reflecting superconducting order, there must also be
a counterpart $F_{pg}$ (discussed in detail elsewhere \cite{NJOP} which 
reflects non-condensed
pairs).
Interestingly, in the weak dissipation
limit the spin transport correlation functions depend only on the
total pairing gap. 
When pg effects are dropped, these equations reduce to their
usual BCS counterparts. 

Any consistent theory of the spin transport must be compared with the
f-sum rule
\begin{equation}
\int_{-\infty}^{\infty}d\omega \omega\chi^{\prime\prime}(\textbf{q},\omega)/\pi=n\textbf{q}^2/m
\end{equation}
where
$\chi \prime \prime (\bf q, \omega) = - \rm{Im} Q_{00}^s / \pi$ and
we have verified from Eq.~(\ref{eq:17}) that this can be proved analytically.

In the weak dissipation limit,
it follows from Eq(\ref{eq:40}) and Ref.\onlinecite{OurBraggPRL}
that the spin conductivity  
$\sigma_s=-\displaystyle{\lim_{\omega\rightarrow 0}}
\displaystyle{\lim_{\textbf{q}\rightarrow 0}}\frac{\omega}{\textbf{q}^2}
\rm{Im} \frac{Q_{00}^s(\textbf{q}, \omega) }{ \pi}$. 
Importantly, it is common to interpret the spin conductivity in terms of an
effective carrier number $\Big[\frac{n}{m}(T) \Big]_{\rm{eff}}$ 
\begin{eqnarray}
\sigma_{s}=
\Big[\frac{n}{m}(T) \Big]_{\rm{eff}}
\tau_s(T)
=\frac{2}{3}\tau_{s}\displaystyle{\sum_{\textbf{p}}}\textbf{v}_{\textbf{p}}^2\Big(-\frac{\partial f}{\partial E_{\textbf{p}}}\Big)
\label{eq:6}
\end{eqnarray}
where $\tau_s = 1/\gamma$ which can be directly
associated with $\frac{1}{\Gamma_{sd}}$, the spin drag relaxation time.
Here
$\textbf{v}_{\textbf{p}}=\partial E_{\textbf{p}}/\partial\textbf{p}$. 
The microscopic Kubo calculation presented here leads to 
an identification between the spin-drag lifetime,
$\tau_{s}$ and the quasi-particle lifetime $\gamma^{-1}$,
since spin
and ``charge" or mass are carried by the same quasi-particles.
This identification was also observed in previous
Kubo calculations of spin diffusion in Helium-3 \cite{ShahzSpin}.
That the inter-conversion between fermions and bosons is the physical
origin of the lifetime leads us to speculate
that $\tau_{s}$ is smallest for the
unitary gas \cite{NJOP} where the number of bosonic pairs and fermions are
roughly comparable. 
This appears
consistent with the findings in
Ref. \onlinecite{ZwierleinSpin}. The spin susceptibility is
$\chi_s^{pg}=2\displaystyle{\sum_{\textbf{p}}}\Big(-\frac{\partial f}{\partial E_{\textbf{p}}}\Big)$.
Alternatively,
from the small $\omega$ and $\textbf{q}$ hydrodynamics, it is seen 
that 
$\chi_s = \sigma_s / D_s$.
\cite{KadanoffMartin}. 
This latter approach is used in recent experiments
\cite{ZwierleinSpin}.

We now turn to these experiments.
In Fig.\ref{fig:1}(a), the values of $\tau_{s}^{-1}$ 
inferred from RF data \cite{OurComparison}
are plotted in units $E_F$ \cite{Ourviscosity}. 
The lifetime increases for lower temperatures 
since inter-conversion
processes cease when non-condensed bosons disappear;
in this way fermions become 
long lived. 
At the lowest $T$ considered, 
this figure shows similar trends to Fig 2 in Ref.\onlinecite{ZwierleinSpin}.
Here one associates
$\Gamma_{sd}$ with
$\tau_{s}^{-1}$.

\begin{figure}
\includegraphics[width=1.9in,clip]
{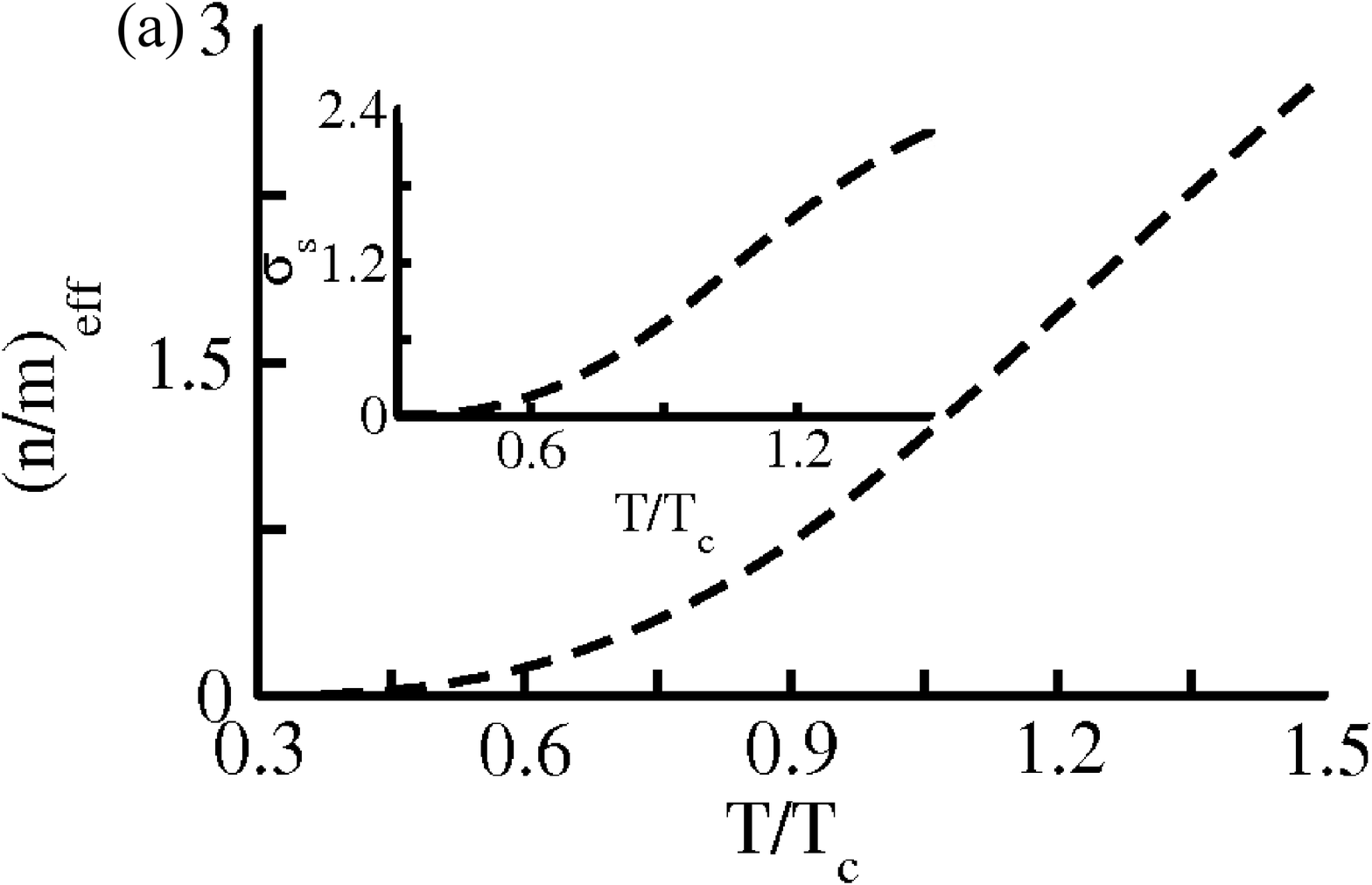} 
\includegraphics[width=1.6in,clip]
{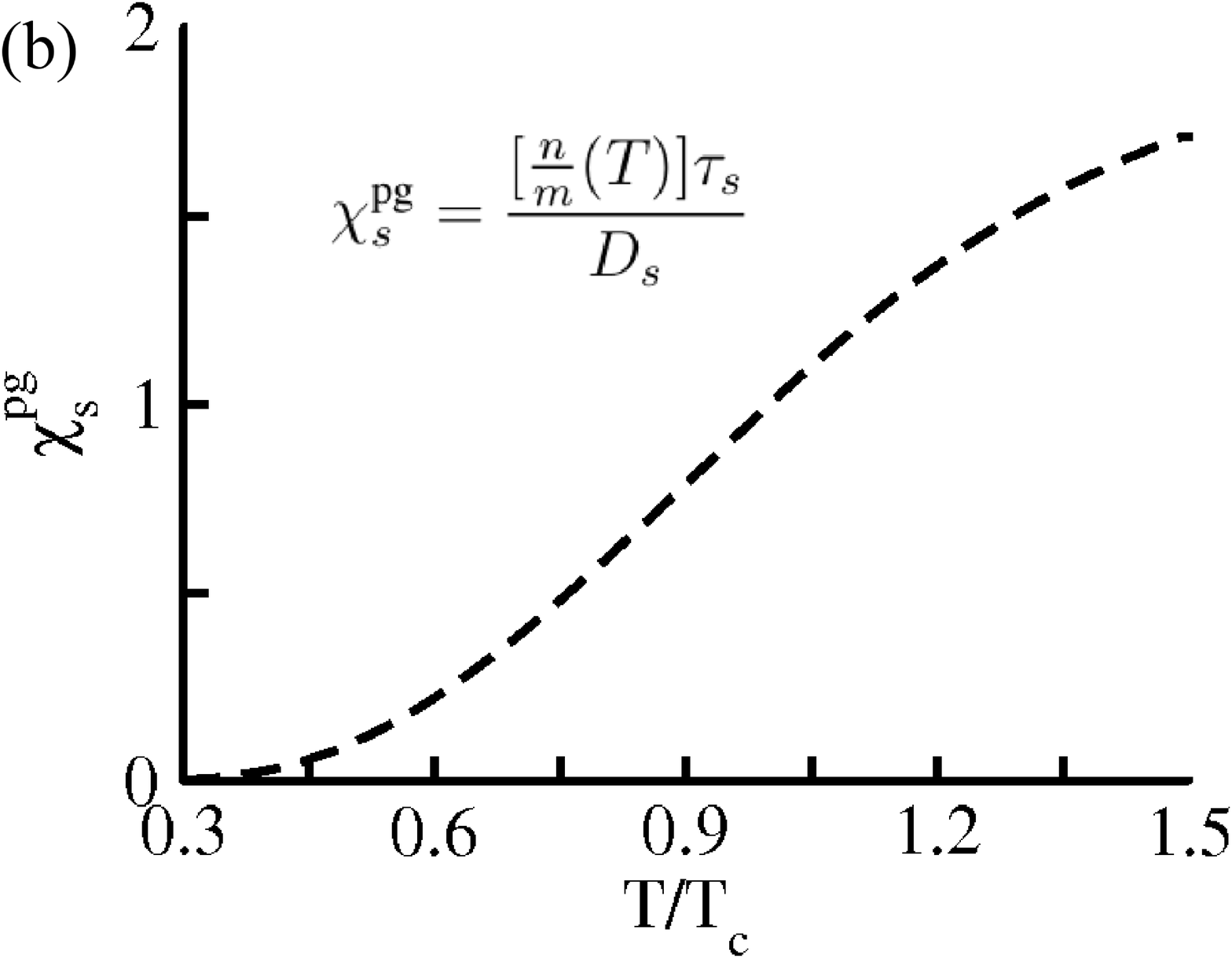}
\caption{(a) The effective carrier number $(n/m)_{\textrm{eff}}$ for a homogenous unitary Fermi gas in our microscopic theory normalized by the value at $T_c$. It vanishes at low temperature because of the pairing gap $\Delta$.
(a) Inset:The spin conductivity $\sigma_s$ for a homogenous unitary Fermi gas in
the microscopic theory normalized by the value at $T_c$.
(b) The spin susceptibility $\chi_s^{pg}$ for a homogenous unitary Fermi gas in
our microscopic theory normalized by the $T_c$ value.
}
\label{fig:2}
\end{figure}
The microscopically computed spin diffusivity $D_s$ is shown in 
Fig.\ref{fig:1}(b). Importantly we find
the spin diffusivity is suppressed at low temperatures reflecting
the suppression in  
the spin conductivity $\sigma_s$ as a result of the reduced number
of carriers.
As in experiment,
$D_s$ is theoretically found to decrease and (at the lower $T$) this figure 
shows similar
trends to Fig. 3 in Ref. \onlinecite{ZwierleinSpin}. We estimate $D_s$ at $T_c$ is $0.61\hbar/m$, which 
when corrected by a factor of 
$~5.3$,
estimated to account for trap effects \cite{ZwierleinSpin}, is in reasonable agreement.
As $T \rightarrow 0$ we cannot rule out an upturn in $D_s$,
depending on $\tau_s(T)$.


The protocol used to deduce the spin susceptibility 
in
Ref.~\onlinecite{ZwierleinSpin} is based on
\begin{equation}
\chi_s^{FL} = \sigma_s/D_s \equiv \frac{n}{m} \tau_s / D_s~~~~\frac{n}{m} = \rm{constant}
\label{eq:7}
\end{equation}
We use the superscript
$FL$ to emphasize that this analysis builds in a Fermi liquid
interpretation by assuming that the carrier number is a constant
in temperature and that no excitation gap is present.
Qualitatively similar, low $T$  experimental behavior is shown in Fig. 4a of Ref. \onlinecite{ZwierleinSpin}.
Nevertheless, this agreement between theory and experiment
should not be assumed to support a Fermi-liquid based
interpretation-- but rather to establish internal consistency.
[If the behavior for $\sigma_s$ is assumed to be Fermi liquid
like, the computed $\chi_s$ will also be so].
As in experiment, 
$\chi_s^{FL}$ increases for low temperature and appears incompatible with a
pairing theory since the spin susceptibility does not vanish for low temperatures.

This will be contrasted with the microscopic 
calculation of the spin susceptibility, based on $\sigma_s$ (Eq.\eqref{eq:6}), within a
pseudogap formulation,
where $(n/m)_{\textrm{eff}}$ is 
strongly temperature dependent. 
Plotted in Fig.\ref{fig:2}(a)
is
this  
effective carrier number $(n/m)_{\textrm{eff}}$ as a function of temperature
while the inset indicates the behavior for $\sigma_s$,
computed from Eq.(\ref{eq:40}). 
Both $\sigma_s$ and
$(n/m)_{\textrm{eff}}$ vanish with decreasing temperature
due to the pairing gap.

Our calculation of
the spin susceptibility $\chi_s^{pg}=\sigma_s/D_s$ is shown in 
Fig.\ref{fig:2}(b). This has the expected temperature dependence 
associated with a pseudogap.
Our key observation here is that one must differentiate between the spin
susceptibility
as computed assuming a constant number of
carriers in $\sigma_s$ and one which is
calculated assuming the number of carriers in $\sigma_s$ increases with $T$
reflecting the presence of pairing.
For the latter case
one recovers a consistent interpretation of
both $\chi_s$ and $D_s$, in contrast to the former Fermi liquid
approach.

We have not, in this paper proved one way or the
other whether there is
strong evidence for a pseudogap associated with the experimental 
results presented in 
\cite{ZwierleinSpin}. What we have established is that the indications
from the spin susceptiblity,
which are used to support Fermi liquid theory are
based on Eq.~(\ref{eq:7}) which is associated with a Fermi liquid like behavior. This
does not establish the absence
of a pseudogap. 
Other claims for this absence, based on thermodynamics 
of a balanced unitary gas \cite{SalomonFL}, can be countered by
noting that the same thermodynamic power fits can be found
for a non-Fermi liquid (for example in BCS theory below $T_c$).
Interpretation of recent experiments \cite{Salomon4} which address the low $T$
normal phase associated with a Fermi gas of
arbitrary imbalance also fall in this category. The
only theoretical phase diagram \cite{ChienPRL} 
(of which we aware)--
addressing where pseudogap and Fermi liquid phases are stable--
predicts this low $T$ phase should be Fermi liquid, as
observed \cite{Salomon4}.

A suggestion for establishing how to rule in or rule out
a pseudogap in the normal phase of the unitary gas is
discussed in Ref.~\onlinecite{ChienFL}. Essential is that
one first confirm the known characteristics of 
the superfluid phase, such as a suppressed spin susceptibility
or entropy, and
then establish that these features vary rather smoothly
persisting somewhat above the transition. These observations
are the counterpart of those first applied to the cuprates
and recent cold gas experiments \cite{KetterleSpin} have suggested new
techniques for measuring these potential spin susceptibility suppressions.

Of great interest is the relation between spin and mass transport.
Because
spin
and ``charge" or mass are carried by the same quasi-particles,
even in the presence of a pseudogap \cite{OurBraggPRL}, spin
and mass transport behave similarly.
The analysis applied in this paper was used to anticipate
that the anomalously low shear viscosity
\cite{Ourviscosity} of the normal state should persist down to
$T \approx 0$, as now observed
\cite{ThomasViscosityScience_online}. 
As in Ref.~\onlinecite{Bruunvisc2} we find
the excitation gap is responsible for this behavior.
It would similarly explain bad metallicity \cite{Emery} in the
pseudogapped high $T_c$ superconductors \cite{NJOP}.
It is striking that experiments from the
high $T_c$ community \cite{AndoRes1} have now tended to focus
on the temperature dependence of the effective number of
carriers (in the presence of a pseudogap) and note that it will
affect transport ``because $n_{\rm{eff}}$ may be changing
with $T$." 
These commonalities highlight the importance of the
ultracold gases as a powerful simulation tool for a wide
class of condensed matter systems.

In summary, in this paper we presented
a theory of spin transport in a
non Fermi liquid (FL) scenario
and have shown that
these experiments do \textit{not} provide evidence against
a pseudogap.
Following the experimental
protocols for extracting the spin susceptibility we find
that
$\chi_s$ appears FL like, but we emphasize
this
reflects a  FL- starting point (presuming $T$-independent
carrier number in $\sigma_s$) in the protocols.
As in experiment, at lower $T$ we find that $D_s$ is suppressed.
Moreover $\sigma_s$ is even more so, as both reflect
the suppressed carrier number due to
pseudogap effects.

This work is supported by NSF-MRSEC Grant
0820054. 
C.C.C. acknowledges the support of the U.S.
Department of Energy through the LANL/LDRD Program.

\bibliography{Review2.bib} 

\end{document}